\documentclass[referee]{raa}
\usepackage{graphicx,times}
\usepackage{natbib}
\usepackage{amssymb,amsmath}
\usepackage{subfigure}
\bibpunct{(}{)}{;}{a}{}{,}

\usepackage[dvipdfm=true]{hyperref}

\begin{document}

   \title{Integrated HI emission in galaxy groups and clusters
$^*$
\footnotetext{\small $*$ Supported .}
}

 \volnopage{ {\bf 2012} Vol.\ {\bf X} No. {\bf XX}, 000--000}
   \setcounter{page}{1}

   \author{Mei Ai\inst{1}, Ming Zhu\inst{1}, Jian Fu\inst{2}
   }

   \institute{ National Astronomical Observatories,20A Datun Road, Chaoyang District, Beijing, China;{\it aimei@nao.cas.cn}, {\it mz@nao.cas.cn}\\
        \and  Key Laboratory for Research in Galaxy and Cosmology, Shanghai Astronomical Observatory, Chinese Academy of Sciences, 80 Nandan Road, Shanghai 200030, China\\
\vs\no
}

\abstract{The integrated HI emission from hierarchical structures such as groups and clusters of galaxies can be detected by FAST at intermediate redshifts. Here we propose to use FAST to study the evolution of the global HI content of clusters and groups over cosmic time by measuring their integrated HI emissions. We use the Virgo cluster as an example to estimate the detection limit of FAST,  and have estimated the  integration time to detect a Virgo type cluster at different redshifts (from z=0.1 to z=1.5). We have also employed  a semi-analytic model (SAM) to simulate  the evolution of HI contents in galaxy clusters.  Our simulations suggest that the HI mass of a Virgo-like cluster could be 2-3 times higher and the physical size could be more than 50\% smaller when redshift increases from z=0.3 to z=1. Thus the integration time could be reduced significantly and gas rich clusters at intermediate redshifts  can be detected by FAST in less than 2 hour of integration time.  For the local universe, we have also used SAM simulations to create mock catalogs of clusters to predict the outcomes from FAST all sky surveys. Comparing with the optically selected catalogs derived by cross matching the galaxy catalogs from the SDSS survey and the ALFALFA survey, we find that the HI mass distribution of the mock catalog with 20 second of integration time agrees well with that of observations. However, the mock catalog with 120 second integration time predicts much more groups and clusters that contains a population of low mass HI galaxies not detected by the ALFALFA survey. Future deep HI blind sky survey with FAST would be able to test such prediction and set constraints to the numerical simulation models. Observational strategy and sample selections for the future FAST observations of galaxy clusters at high redshifts are also discussed.
}

   \authorrunning{M.Ai et al. }            
   \titlerunning{Integrated HI emission in galaxy groups and clusters}  
   \maketitle

%
\section{Introduction}           
\label{sect:intro}

In the hierarchical scheme of structure formation, massive clusters of galaxies are the last objects to form, becoming prominent only at $z < 1$.  Our understanding of galaxy evolution within the cluster formation scenario is dominated by two principal observational cornerstones: the strong signature of morphological segregation exhibited within rich clusters today and the Butcher-Oemler effect, i.e., the increase in the blue cluster population with redshift. Observational tests of our understanding of the details of how environment (cluster formation) affects individual galaxies (morphology, gas content, star formation rate) are critically needed to validate the predictions of numerical simulations within the cluster formation scenario. In the local universe, the relative HI content serves as a useful comparative indicator of the future star formation potential of a galaxy. There have been many observational studies of nearby galaxy clusters,  mostly in optical and X-ray wavelengths. However, study of HI selected galaxies of cluster is rare, e,g, \citealt{Freudling1989}. Its simple physics makes the HI line transition an attractive potential tracer of galaxy evolution over cosmic time. However, because of the inherent weakness of the HI line and the contamination of the spectrum by terrestrial radio frequency interference (RFI), studies of redshift HI emission from individual galaxies have only recently been possible up to 0.25-0.37 (\citealt*{Catinella2008}; \citealt{Fernandez2016}). Beyond that, the study of HI emission from individual objects requires the resolution and collecting area of the future Square Kilometer Array (SKA).

Nowadays there are many studies on HI properties on the local Universe cluster (\citealt*{Giovanelli1985}, \citealt{Solanes2001}, \citealt{Chung2009}, \citealt{Wolfinger2013}) or group scales (\citealt{Stevens2004}, \citealt{Sengupta2006}), or on the HI properties between galaxies in group or clusters (\citealt{English2010}).  Most of these researches focused on the influence of cluster environment on the HI content in galaxies. Some local spiral-rich clusters, such as Virgo, Herclues, Ursa, Abell 3128, have been observed by radio telescopes, and these investigations showed that most HI gas resides in spiral galaxies, elliptical galaxies contains limited amount of HI gas. Young clusters consisting of more spiral galaxies contain more HI than old clusters with more elliptical galaxies. Some mechanisms such as ram-pressure \citep*{Gunn1972}, viscous stripping \citep{Nulsen1982}, tidal interaction \citep{Merritt1983} could contribute to the HI decreasing  process during the evolution of clusters/groups. The core region of a cluster is usually hot and most HI content resides at the edge of the cluster, but recent studies show that galaxies in a cluster begin to loose their HI gas at intermediate distance to the cluster center \citep{Chung2009}. The relative importance of HI deficiency mechanisms in clusters/groups in the local Universe is still debated.

Now that the contruction of the  Five-hundred-meter Aperture Spherical radio Telescope(FAST) has been completed , we propose to use FAST to study the evolution of the global HI content of clusters and groups over cosmic
time by measuring the integrated HI emission from clusters. The sensitivity of FAST is high enough to detect  HI emission from galaxy groups and clusters at intermediate redshifts. In this paper we discuss
the observational strategy and sample selections for future FAST deep HI surveys. We also use numerical simulations to predict the global HI contents
on cluster scale at different redshifts.  In all the calculations we adopt the $\Lambda$DCM cosmology parameters: $H_{\rm 0}$ = 70 $\rm kms^{-1}Mpc^{-1}$, $\Omega_{\Lambda}$=0.7, $\Omega_{m}$=0.3 unless otherwise stated.

\section{FAST observations of integrated HI emission in galaxy clusters/groups}
\label{sect:Obs}

\subsection{Technical specifications of FAST}

Construction of FAST major structure was completed in September 2016. This telescope  has an aperture of 500m and an illuminated aperture of 300m in diameter. It is located in southern China, Guizhou province, with an latitude of $25.65^{\circ}$.  FAST can reach a zenith angle of $40^{\circ}$, so the sky coverage of declination would be between $-14^{\circ}$ and $65^{\circ}$. The detailed technical specifications of FAST can be found in the paper of \cite*{Nan2011} and an updated status is in
\cite*{Li2016}.
There are 7 sets of receivers  being developed for FAST. The major ones used for HI observations are the 19-beam feed-horn array receiver covering the frequency range of 1.05-1.45 GHz with $T_{\rm sys}$=25K. For galaxies at redshift higher than 0.35 and up to 1.5, a single-beam receiver covering the frequency range of 560 MHz-1.12 GHz can be used. This receiver has a better system temperature of about 20K, which is ideal for high redshift HI observations.

FAST has very high sensitivity, but its resolution is relatively low comparing to interferometers. One of the most important science goals of FAST is to investigate the HI distribution in the Universe. As the main beam of the FAST is relatively large (2.95'), there could be more than one galaxies in the FAST beam, especially in distant clusters and close groups. Thus in this paper we will investigate detecting the total HI content in groups or clusters, without distinguishing individual galaxy members. Many groups and clusters are gravitationally bound structures, thus our study can reveal the large scale HI structure of the local Universe.

\subsection{Science goal:  evolution of HI contents in galaxy groups/clusters}

The major goal for high-z HI observations is to study the  evolution of HI gas contents.
FAST is expected to detect HI on cluster scales at intermediate redshifts. This allows us to study the evolutionary effect of HI content in cluster at different redshifts.
Butcher and Oemler \citep{Butcher1978}  pointed out that there are a substantial population of blue galaxies in clusters at $z \geq 0.4$ , while nearby clusters are dominated by elliptical and lenticular galaxies. The dramatic transformation of cluster galaxies happened during redshift between $0 \sim 0.4 $, a look back time of 4-5 billion years. However, HI observations at present are limited to low redshift Universe. The highest redshift of  the ALFA extragalactic HI survey  (ALFALFA) using the Arecibo 305 meter telescope  is 0.06 (Giovanelli et al. 2005 ).  The HIGHz Arecibo survey detected 39 galaxies at $0.16<z<0.25$ with $M_{HI} = (3-8) \times 10^{10} M_{\odot}$ \citep*{Catinella2015}. About 160 galaxies in two cluster regions at $z \sim 0.2$ are detected by the Blind Ultra Deep HI Environmental Survey(BUDHIES) with the Westerbork Synthesis Radio Telescope(WSRT), with HI mass ranging from $5 \times 10^{9} M_{\odot}$ to $4 \times 10^{10} M_{\odot}$ (\citealt{Jaffe2013}, \citealt{Verheijen2007}). One galaxy (J100054.83+023126.2) at $z=0.376$ is detected by the COSMOS HI Large Extragalactic Survey (CHILES, \citealt{Fernandez2016}) which is the highest redshift galaxy so far observed at the HI line. The HI mass of this galaxy is $2.9 \times 10^{10} M_{\odot}$. The number of detected HI galaxy clusters at present is too small for a statistical study of the HI gas content on cluster scales.

Deep HI observations with FAST should be able to detect cluster scale HI gas at redshifts high enough to study the B-O effect. Evolution of the  HI gas content in clusters as a function of redshift would be an important science goal for future FAST HI surveys. This will not only help to understand the relative importance of the mechanisms for HI deficiency in nearby clusters, but can also shed light on the global evolution of galaxy clusters.

\subsection{FAST detection limit for gas rich clusters: a case study of Virgo}

Virgo is one of the nearest rich galaxy clusters. It has a complex structure comprised of at least three distinct clouds, suggesting that Virgo cluster is on a relatively early stage of evolution. Unlike the Coma cluster where there is almost no HI detected, there are still a large amount of HI rich galaxies in the Virgo cluster. Here we use Virgo as an example to evaluate the ability of detecting the integrated HI emission at high redshifts with FAST. The Virgo cluster region covering 130 $\rm deg^2$ over $12^h04^m<$\rm R.A.$<12^h44^m$ , $3^{\circ} <$\rm decl$<16^{\circ}$ , $\rm 0<V\le3000kms^{-1}$. The exact boundary of the Virgo cluster has limited effect on our estimates.

We extract the HI sources in this region from the ALFALFA 70 ($\alpha. 70$) catalog \citep{Haynes2011}, which contains 70\% of the survey data. We include both  ALFALFA "code 1"  and  "code 2" sources, which have a signal-to-noise ratio S/N $>$ 6.5, and 4.5 $<$ S/N $<$ 6.5, respectively.  They are likely to be real because nearly all of them have a known optical counterpart at the redshift of the HI source \citep{Haynes2011}. There are 350 HI sources from ALFALFA in the Virgo region. As shown on Fig. 1, HI sources are plot as blue circles with radius proportional to their HI mass which is varying between $10^{6}M_{\odot}$  and $10^{10.07}M_{\odot}$. To estimate the integration time of FAST for observing such object, we treat the whole Virgo cluster as one "synthesis" HI galaxy. We put the cluster at different redshifts and calculate the integrated HI fluxes. The luminosity distances of different redshifts are listed in Table 1. FAST beam size is 2.95$'$ at z=0 and varies with redshifts as $2.95'(1+z)$. The physical beam size is calculated from the angular size distance $D_{\rm A}$ multiplied by the radian FAST beam size at the corresponding redshift where $D_{\rm A} = \frac{1}{1+z}\frac{c}{H_0}\int_0^z \frac{dz}{\sqrt{\Omega_{\Lambda}+\Omega_{m}(1+z)^{3}}}$.  According to the theory of dark matter halo growth \citep*{White1991}, the radius of a virialized halo at redshift z follows the relation  $ r_{\rm vir}=0.1H_0^{-1}(1+z)^{-3/2}v_c$. Hence we can assume that the size of Virgo Cluster at redshift z is $r_0(1+z)^{-3/2}$, where $r_0$=1.95 Mpc is the Virgo cluster radius at z=0. So the cluster size should be smaller at higher redshift. As z increases, the FAST beam would cover larger physical area of the Virgo cluster, and more galaxies will contribute to the observed HI fluxes. At z=0.7, the cluster diameter would be 1.92 Mpc which is less than the physical beam size at this redshift and all the Virgo HI galaxies are incorporated in the beam. It should be noted that here we only consider the size evolution of the cluster and did not consider the evolution of the HI gas mass. In Section 3.2 we provide a more accurate estimate of the cluster size and HI mass based on cosmology simulations.

On Fig. 1 we overlay the beam circles (black dashed circles) at different redshifts on the HI sources distributed on physical scales. The smallest black dashed circle corresponds to the physical beam size where we put the cluster at redshift of 0.5 and the largest one corresponds to redshift of 0.7. The redshift interval of the black dashed beam circles is 0.1. The blue filled circles are the HI galaxies reside in the Virgo cluster region that we extracted from the $\alpha. 70$ catalog. The radius of the blue filled circles are proportional to the HI mass.

\begin{figure}
  \centering
   \includegraphics[width=143mm,height=142mm]{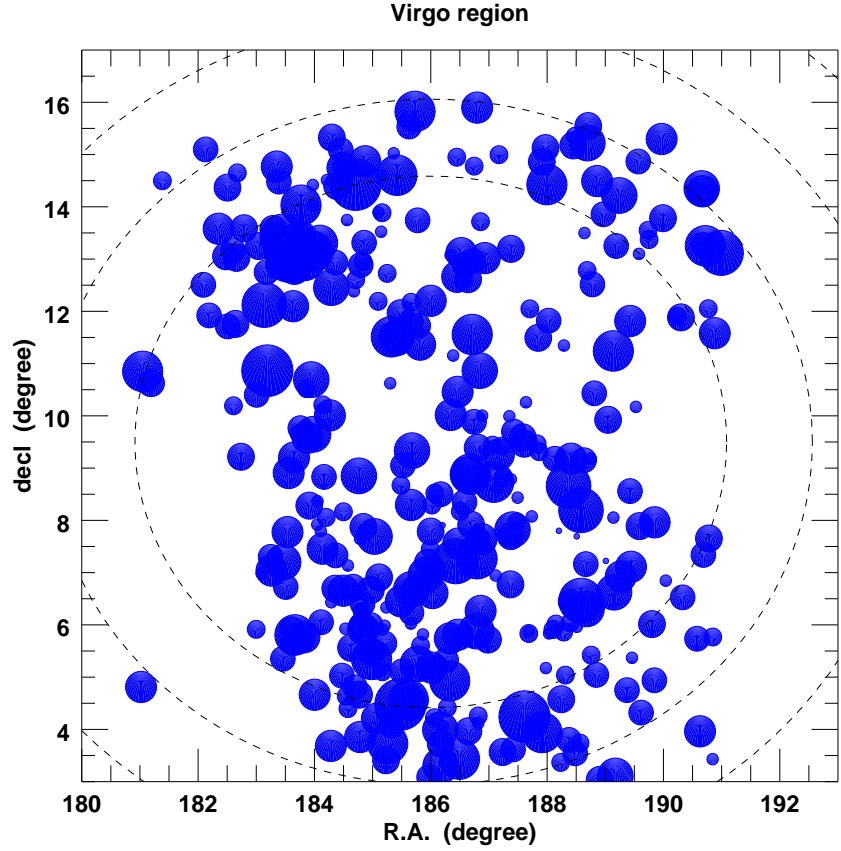}
   \caption{{\small HI sources within the Virgo cluster region. Blue circles represent the HI galaxies, with their areas proportional to their HI masses.  The black circles indicate the physical area covered by the FAST beam at a distance of z=0.5 to z=0.7. The redshift interval is 0.1} }
\end{figure}

To estimate the FAST detection limit, we add the HI mass of all sources covered by the beam circle, and use the co-added mass to estimate the HI flux and the required integration time of FAST:
\begin{equation}
S_{\rm peak}=\frac{M_{\rm HI}}{2.36\times{10^5}\cdot{D^2_{\rm Mpc}}\cdot{\omega_{\rm km/s}}}
\end{equation}

\begin{equation}
S_{\rm noise}=\frac{k}{\frac{A}{T_{\rm sys}}\sqrt{\tau\cdot\delta\nu}}
\end{equation} and assuming

\begin{equation}
S_{\rm noise}=\frac{S_{\rm peak}}{5}
\end{equation} so

 \begin{equation}
\tau=(\frac{k}{S_{\rm noise}\cdot\frac{A}{T_{\rm sys}}})^2\cdot\frac{1}{\delta\nu}
\end{equation}

where $\omega_{\rm km/s}$ is the estimated velocity widths which we assumed according to different redshifts, $D_{\rm Mpc}$ is the redshift where we put the Virgo cluster, $\tau$ is integration time, $\frac{\rm A}{T_{\rm sys}}$ is the sensitivity of FAST which is taken as 2000$\rm m^2/K$ \citep{Nan2011}, k is the Boltzmann constant, $\delta\nu$ is frequency resolution which is assumed to be 142 kHz, corresponding to a velocity resolution of 30 km/s. The results are shown in Table 1. In Table 1, z is the redshift where we put the cluster, $M_{\rm HI}$ is HI mass within the FAST beam, $S_{\rm peak}$ is the corresponding peak flux, $\tau$ is the integration time, $\omega$ is the velocity width of the assumed HI source. For redshift from 0.1 to 0.4 we centered the beam at R.A.=183.5, decl=13 which is the local concentration that contains more HI galaxies than in the center. The Beam-size in column 7 is the physical diameter of the FAST beam when observed at different redshifts. $N_{\rm beam}$ of column 8 is the number of HI galaxies in the FAST beam, the luminosity distance of each redshift is in column 9. The velocity width, $\omega_{\rm km/s}$, is determined based on the velocity dispersion of galaxy group/cluster catalog extracted from the SDSS and 2MASS surveys. Its value is similar to the group velocity dispersion when $N_{\rm beam}$ is close to the group number of member galaxies.

In the above analysis we do not consider evolutionary effect of the HI content at higher redshift. \cite{Rhee2016} conclude that there is no significant evolution in cosmic HI mass
density from $z=0 \sim 0.4$. The weighted mean average of the HI mass density from all 21-cm measurements at redshifts $z<0.4$ is $ (0.35 \pm 0.01) \times 10^{-3}$. However, the Butcher-Oemler effect(\citealt{Butcher1978}; \citealt{Butcher1984}) pointed out that clusters at $z\ge0.4$ have a substantial population of blue galaxies implying more HI rich galaxies, while the nearby rich clusters are very deficient in HI gas.   As shown in Section 3.2, numerical simulations predict that the HI mass of a cluster could be about 2-3 times larger when  z increase from 0.3 to 0.8.  Thus the integration time for high z gas rich cluster can be 1/4 - 1/9 time less than that listed in Table 1.  If this is the case, a cluster
at z=0.8 can be detected by FAST in about 2 hour of integration time.

\begin{table}
\bc
\begin{minipage}[]{100mm}
\caption[]{The observing parameters of FAST when using the Virgo Cluster as a sample\label{tab1}}\end{minipage}
\setlength{\tabcolsep}{1pt}
\small
 \begin{tabular}{ccccccccccccc}
  \hline\noalign{\smallskip}
Redshift& $M_{HI}$  & $S_{peak}$  & $\tau$  & $w$   &  Beam position & Beam size & $N_{\rm beam}$ &Luminosity Distance\\
 & $10^{10}M_{\odot}$& $10^{-2}$mJy&  minit&    km/s&  &Mpc&&Mpc\\
  \hline\noalign{\smallskip}
   0.1&  1.30&52.0&   5& 500&(183.5,13) &0.36 &17&460.3\\
   0.2&  2.26&19.9&   35& 500&(183.5,13)&0.70 &38&980.1\\
   0.3&  4.20&14.8&   64& 500&(183.5,13)&1.02 &69&1552.7\\
   0.4&  5.35&9.6&    151& 500&(183.5,13)&1.33 &99&2172.1\\
   0.5&  9.75&7.4&    258& 700&(186,9.5)&1.62 &249&2832.8\\
   0.6&  15.69&7.6&   240& 700&(186,9.5)&1.89 &335&3530.1\\
   0.7&  16.20&5.4&   478& 700&(186,9.5)&2.15 &350&4259.5\\
   0.8&  16.20&3.9&   920& 700&(186,9.5) &2.39&350&5017.5\\
   0.9&  16.20&2.9&   1645&700&(186,9.5)&2.62&350&5801.0\\
   1.0&  16.20&2.3&   2768&800&(186,9.5)&2.83&350&6607.2\\
   1.1&  16.20&1.8&   4436&800&(186,9.5)&3.04&350&7433.8\\
   1.2&  16.20&1.4&   6824&800&(186,9.5)&3.23&350&8278.8\\
   1.3&  16.20&1.2&   10141&800&(186,9.5)&3.41&350&9140.5\\
   1.4&  16.20&1.0&   14630&800&(186,9.5)&3.58&350&10017.5\\
   1.5&  16.20&0.8&   20571&800&(186,9.5)&3.74&350&10908.4\\
  \noalign{\smallskip}\hline
\end{tabular}
\ec
\tablecomments{0.86\textwidth}{Column 1 is the redshift where we put the cluster, column 2 is HI mass within the FAST beam, column 3 is the peak flux, column 4 is integration time, column 5 is the velocity width of the assumed HI source,column 6 is the position of beam center, column 7 is physical size of FAST beam when it is observing at different redshift, column 8 is the number of HI galaxies in FAST beam, column 9 is the luminosity distance.}
\end{table}

\subsection{Detecting optical selected clusters with FAST}

To further evaluate the feasibility of detecting high reshift galaxies based on optically selected sample (\citealt*{Wen2012},\citealt*{Wen2015}, WHL cluster catalog),  we use the WHL optical selected cluster catalogue  to estimate the HI mass of higher redshift clusters. We select four clusters at redshift around 0.1, 0.2, 0.3 and 0.4. These four clusters represent the richest clusters at each redshift. We retrieve the g and r band absolute magnitude from SDSS Casjobs (absMagG and absMagR from Photoz table, k corrected to z=0) for the WHL cluster member galaxies using their position (R.A. and decl).  Then we use the calculated r band luminosity and g-r color to compute each member galaxy's HI mass using the following equations(\citealt{Bell2003}, \citealt{Zhang2009}):

\begin{equation}
\log (M_*/L)=-0.306+1.097(g-r)
\end{equation}
and
\begin{equation}
 \log(G/M_*)=1.09431-3.08207(g-r)
\end{equation}

where L, $M_*$ is the r band luminosity, stellar mass of the member galaxy, respectively, G is the calculated HI mass.

All these four clusters have the radius of about $\sim 2$Mpc which is larger than the FAST beam size at the corresponding redshift. We calculate the HI mass that covered by the beam at each redshift. In each cluster we consider only the galaxies covered by the FAST beam. We add those galaxies' HI mass and use this mass using Eq (1) to (4) to calculate the integration time needed for FAST to detect such cluster. The results are shown in Table 2. Column 1 is the WHL cluster name, column 2 is cluster photometric redshift, column 3 is cluster radius, column 4 is the number of member galaxies within $r_{200}$, column 5 is the radius of the FAST beam at different redshifts, column 6 is the number of member galaxies that covered by the beam, column 7 is the HI mass of the cluster, column 8 is the HI mass covered by the FAST beam, column 9 is the velocity widths, column 10 is peak flux of the cluster HI sources, column 11 is the estimated integration time of FAST. From Table 2 we can see that all these 4 clusters can be detected by FAST in less than 40 minutes.  However, the HI masses estimated from Eq (5) and Eq (6) carry large uncertainties. The scatters of Eq (5) and (6) is 0.1 dex and 0.35 dex, respectively.  Hence the integration time could be several times longer than that listed in Table 2.  In the worst situation, the integration time for detecting the WHL cluster at z=0.4 would be more than 4 hours. Thus target selection is crucial for detecting high redshift clusters.

\begin{table}
\caption[]{Parameters of optical WHL selected clusters\label{tab1}}
 \begin{tabular}{ccccccccccccc}
  \hline\noalign{\smallskip}
Clustername&$z_{\rm ph}$&$r_{200}$ &$N_{\rm 200}$ & Half beam size &$N_{\rm gal}$ &$M_{\rm HI}$ &$M_{\rm HI}$ in beam  & $\omega_{\rm km/s}$ & $S_{\rm peak}$  & $\tau$  \\
  && Mpc& &Mpc&& $\rm 10^{11}M_{\odot}$ &$\rm 10^{10}M_{\odot}$ & $\rm km\;s^{-1}$&$\rm mJy$& min \\
  \hline\noalign{\smallskip}
 J004629.3+202805& 0.1040 & 1.88 &121&0.18&5    &   7.78 &1.37    & 1000& 0.27 &   19\\
 J092200.5+515521& 0.2042 & 1.98 &128&0.35&13   &   11.34&7.24    & 1000& 0.31 &   14\\
 J233739.7+001617& 0.3042 & 1.93 &103&0.51&26   &   10.25&26.46   & 1500& 0.30 &   15\\
 J081025.7+181724& 0.4123 & 1.71 &80 &0.67&61   &   7.83 &31.78   & 1500& 0.19 &   40\\
 \noalign{\smallskip}\hline
\end{tabular}
\end{table}

\section{Theoretical predictions}

\subsection{HI contents in nearby groups and clusters}

In order to understand the integrated HI contents in high redshift clusters, we need to first study the HI properties in the groups and clusters at low redshifts. In the hierarchical galaxy formation model, disk galaxies form first and then fall into bigger halos. By merging, dark matter halos grow over cosmic time. Galaxies tend to appear in groups, or clusters. Groups have fewer galaxy members than clusters, but can grow into clusters when more members are accumulated. Thus we include both galaxy groups and clusters in our mock catalog.

\subsubsection{HI data in optically selected galaxy groups and clusters}

Before creating a mock catalog of groups/clusters, we first need to choose
a control sample of galaxy clusters to make comparison with the simulation results.  \cite{Berlind2006} have derived a cluster catalog  from the SDSS  DR7 survey using an optimized algorithm of the friends of friends (FoF) method (available online \url{http://lss.phy.vanderbilt.edu/groups/dr7/}). There are three galaxy samples created with different absolute magnitude limits and redshift ranges. Each one is complete to within the stated limits. Given that the highest redshift of the HI sources in ALFALFA  is around 0.06 \citep{Giovanelli2005} and the luminous HI source corresponds to less luminous optical galaxies, we choose the faintest Mr18 group sample whose r band absolute magnitude is down to $M_r=-18$. Then we cross-match the SDSS Mr18 cluster member galaxy catalog with the ALFALFA 70\% ($\alpha.70$) HI source catalog(see the ALFALFA 40\% description \citealt{Haynes2011}). The $\alpha.70$  catalog which contains 70\% of all the ALFALFA data is available on the web site \url{http://egg.astro.cornell.edu/alfalfa/data/}. In this catalog, the detected HI sources with the SDSS optical counterparts (OCs) available have been identified and their (RA, DEC) coordinates are listed. We cross match the Mr18 cluster member galaxies with the $\alpha.70$ HI sources using the software package TOPCAT, and the angular difference between SDSS member galaxies and $\alpha.70$ HI OC's coordinates is 5$''$. There are 1443 SDSS clusters (9868 member galaxies) in the selected region with N$\ge$ 3, where N is the number of member galaxies in a cluster. However, only 2134 $\alpha.70$ HI sources are cross matched with SDSS member galaxies within 977 clusters. Thus the total detection rate is  only 2134/9868=  22\% for the optically selected sample. High sensitivity observations with FAST can greatly improve the detection rate of HI galaxies in groups which is crucial for studying the evolution of HI contents in groups and clusters.

The final control sample of HI groups/clusters are the cross matched SDSS-ALFALFA clusters which fall in the ALFALFA Spring Sky region with $114 <$\rm R.A.$ <248$, $0^{\circ}<\rm Dec<30^{\circ}$. The redshift range of the SDSS clusters is $0.02<\rm z<0.042$.

\subsubsection{A mock catalog of groups and clusters}

We adopted the same method in \citealt{Duffy2012} to create the galaxy catalog by the semi-analytic
model (SAM) of \citealt{Croton2006}, which is based on the underlying dark-mater-only MILLENNIUM
SIMULATION \citealt{Springel2005}. This sample of galaxies can accurately recreate the observed stellar mass
function with both supernovae feedbacks and feedbacks at the high-mass end due to active galactic nuclei.
The cosmology used in the MILLENNIUM SIMULATION is (M = 0.25,c = 0.75,b = 0.045, $\sigma_8$ = 0.9 and
Hubble parameter h = 0.73). Using the sensitivity parameters of FAST with the 19 beam array receiver, we make use of the CLOUD-based web application of Theoretical Astrophysical Observatory (TAO) which generates mock catalogs from different cosmological simulations and galaxy models in the form of a light cone to create an all-sky galaxy catalog extending to z$<$ 0.35 for the FAST HI survey. Two sets of mock HI catalogs were generated with the help of Alan Duffy (private communication). They are simulated with an integration time of 20 second and 120 second with FAST, corresponding to a sensitivity of 0.7mJy and 0.3 mJy according to Eq (1) and (2).
A shallow survey with FAST with 20s of integration per point would reach a sensitivity of 0.7 mJy, which is slightly deeper than
that of the ALFALFA survey.  Deeper surveys are required to detect more clusters at higher redshifts.

We run the FoF program to extract galaxy clusters from the mock HI catalog. The cluster's total HI mass is derived by summing up all the galaxies' HI masses in that cluster. We are aware of the fact that it is not easy to choose a set of appropriate linking length for a HI galaxy catalog because HI galaxies tend to reside in the outer region of a cluster and some galaxies in the cluster can not be detected by HI. To evaluate this problem, we first choose the linking length used by the SDSS Mr18 catalog \citep{Berlind2006} (0.69Mpc, 259$\rm km\;s^{-1}$). Then we try different sets of linking lengths (0.6Mpc, 400$\rm km\;s^{-1}$; 0.7Mpc, 500$\rm km\;s^{-1}$; 0.3Mpc, 500$\rm km\;s^{-1}$) and found the extracted mock clusters are similar. Therefore, we choose the ones that used by the SDSS Mr18 catalog (0.69Mpc, 259$\rm km\;s^{-1}$) to extract HI mock clusters from the two sets of mock HI catalogs. The largest redshift of the clusters can be extracted from the mock catalog galaxies is 0.08 because there are only a few HI galaxies at farther distance and the distribution of those galaxies are too diffuse to extract clusters. There are 3985 mock clusters with 23447 HI mock sources from the mock HI catalog with the integration time of 120s and 1359 clusters with 6034 sources from the 20s integration time catalog in the sky region identical to the control sample.

\subsubsection{Comparison between the mock catalog and the SDSS catalog}

\begin{figure}
   \includegraphics[width=143mm,height=132mm]{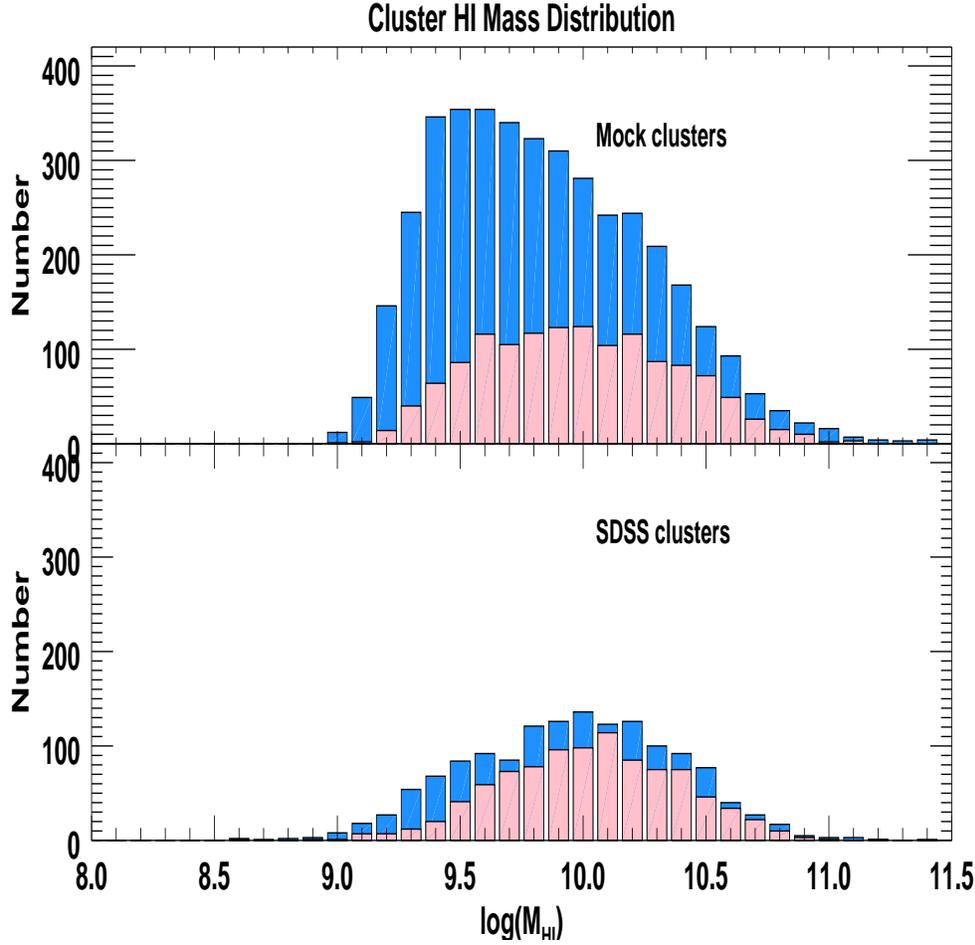}
   \caption{The cluster HI mass distribution of two samples of clusters based on SDSS and mock galaxy catalogs. In the upper panel, the pink and blue bars represent the HI mass distribution of mock cluster that are extracted from the 20s and 120s integration time catalog. In the lower panel, the pink bars show the distribution of the 977 SDSS clusters whose HI masses are derived from cross matching with the $\alpha.70$ catalog, and the blue bars are the distribution of all the 1443 clusters whose HI mass are derived from the cross matched HI sources plus the HI mass of the un-matched member galaxies estimated by galaxy color. See the text for details.}
\end{figure}

Fig. 2 shows the comparion between the SDSS selected clusters and the mock clusters, with number of member galaxies N greater than 3. The pink and blue bars on the upper panel  are the distribution of the mock clusters as a function of cluster HI mass,  which are the simulated outputs with 20s and 120s integration time using FAST. In the lower panel of Fig.2, the pink bars show the distribution of the SDSS clusters vs.HI masses which is obtained in Section 3.1.1. We can see that the 20s mock catalog in general looks quite similar to the SDSS-ALFALFA cross match sample, suggesting that numerical simulations successfully reproduce the observed universe. However, the mock catalogs (both the 20s and 120s ones) appear to have more groups at the low mass end.  This could be due to the fact that the 20s mock catalog  has a rms of 0.7 mJy which is deeper than that of the ALFALFA survey (1.3 mJy), and thus contains more low mass HI galaxies.  As mentioned in section 3.1.1,  the ALFALFA HI detection rate of the SDSS member galaxies is only 22\%.
Thus we need to get estimates the HI mass for a large number of SDSS galaxies not detected by ALFALFA.  To do so,  we use the relationship between HI gas content and optical color g-r as described in Section 2.3.
The g-r color and absolute magnitude of each member galaxy are listed in the SDSS Mr18 cluster catalog, thus we can estimate the
HI mass for the un-matched SDSS member galaxies using Eq (5) and (6).  In this way, we obtained more clusters (1443, instead of 977) and recovered
more HI mass for each cluster. The results are shown as the blue bars on the lower panel of Fig.2.  The distribution of this blue histrogram agrees much
better with that of the 20s mock catalog.  This suggests that 20s integration with FAST can achieve much higher detection rate for nearby groups and clusters.

\begin{figure}
   \includegraphics[width=143mm,height=132mm]{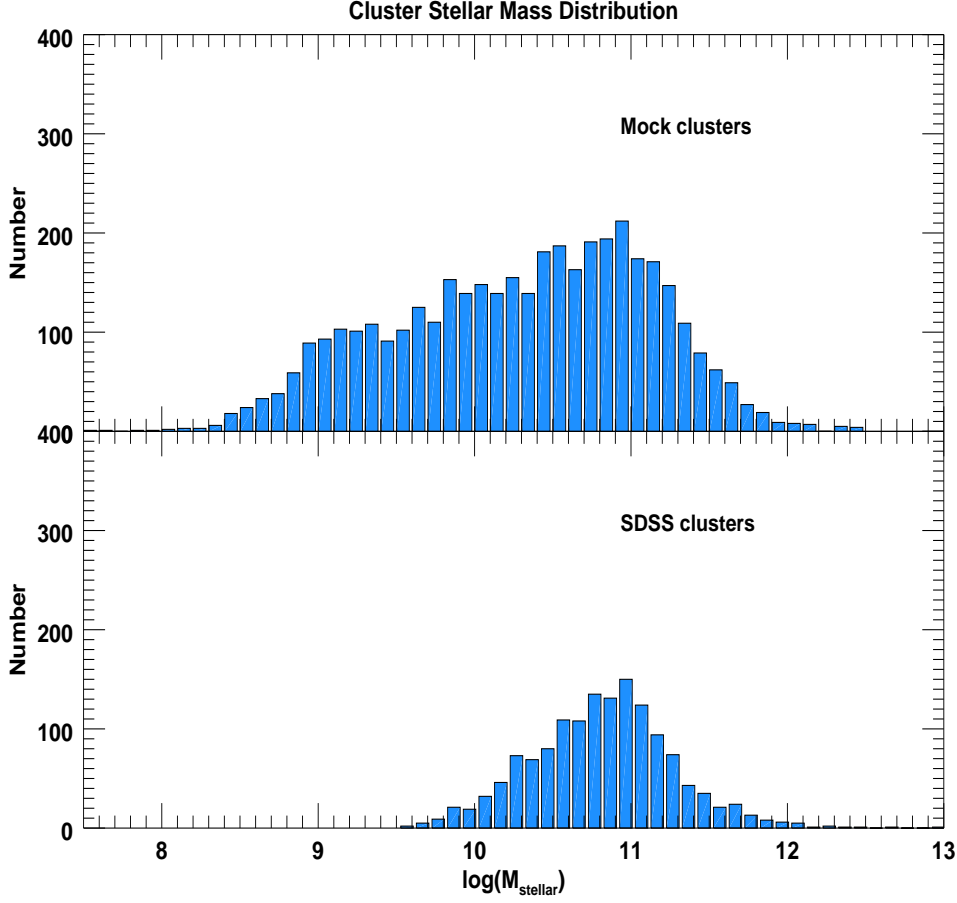}
   \caption{Comparison of the stellar mass distribution of the SDSS clusters and the 120s mock clusters. The upper panel is the cluster stellar mass distribution of the 120s mock clusters and the lower panel is the distribution of the SDSS clusters.}
\end{figure}

Longer integration time are desire for detecting more  member galaxies at higher redshifts. Hence we have also generated a mock catalog with 120 seconds of integration. It is remarkable that numerical simulations predict significantly more clusters in the  120s mock catalog, as shown in the blue histrogram on the upper panel of Fig.2.  The 120s mock catalog has much more HI low mass groups and clusters, comparing to the control sample and the 20s mock catalog. A plausible explanation is that SAM simulations predict a large amount of low mass HI galaxies which are optically too faint
to be included in the SDSS Mr. 18 catalog  (sources fainter than -18 mag are not included). To evaluate such effect,  we compute the minimum detectable HI mass at different distances using Eq (5) of \cite{Giovanelli2005} assuming  a velocity width of $\rm 200\;km\;s^{-1}$ for a galaxy, but use $3\sigma$ here instead of $6\sigma$ in that equation.  The 3 $\sigma$ detection limit of HI mass in ALFALFA with an integration time of 40 seconds is $2.78\times10^8 M_{\odot}$  at z=0.02 and $1.11\times10^9 M_{\odot}$ at z=0.04.  Similarly, assuming a frequency resolution of 142kHz (velocity width of $\rm 200\;km\;s^{-1}$), the HI mass that could be detected by FAST in 120s of integration time is $2.33\times10^8 M_{\odot}$ at $z=0.04$ ($5.83\times10^7 M_{\odot}$ at $z$=0.02).  Hence galaxies with HI mass in the range $2.78\times10^8 M_{\odot} < M_{HI} < 1.11\times10^9 M_{\odot}$ at z=0.04, for example, would be included in the 120s mock catalog, but not in the SDSS-ALFALFA catalog.

We have further examined the stellar mass of the galaxies in the 120s mock catalog and found that the SAM simulations produce surprisingly large amount of dwarf galaxies. Such effect is similar to the "missing satellite" problem. Fig.3 compares the stellar mass distributions of the mock and SDSS clusters. The upper panel of Fig.3 shows the cluster stellar mass distribution of the 120s mock clusters and the lower panel shows the cluster stellar mass distribution of 1443 SDSS clusters. The stellar masses of the mock galaxies are generated from the SAM simulation. The stellar masses of the SDSS galaxies are calculated using Eq (5) with r band absolute magnitude and g-r color which are listed in the SDSS member galaxy catalog. The cluster stellar mass is the sum of all the member galaxy stellar masses in that cluster. It is obvious that there are too many low stellar mass galaxy groups in the 120s mock catalog comparing to the observed SDSS clusters. Many of the dwarf galaxies have a $M_{\rm HI}$/$M_*$ ratio higher than 1. Hence the 120s mock catalog predicts large amounts of low HI mass groups/clusters. To test such prediction,  we will need to compare simulation data with observed data at both HI and optical wavelengths, with sensitive high enough to recover all the member galaxies within the virial radius of a group or cluster. Future high sensitivity observations with FAST, as well as with optical telescopes would be crucial for such studies and tackling the "missing satellite" problem.

\subsection{Detecting high redshift clusters with FAST}

In this part we discuss the predictions of the HI gas component in high redshift galaxy groups or clusters, which can be observed by FAST. The high-z mock catalogues of HI gas are from the outputs of L-Galaxies SAMs (\citealt{Guo2011}, \citealt{Guo2013}) based on both Millennium and Millennium II halos. We adopt the model versions by \cite{Fu2013} and \cite{Luo2016}, which offer the results of both H$_2$ and HI gas in galaxies. In the models, two prescriptions are adopted to calculate the transition between H$_2$ and HI gas. In KMT prescription \citep*{Krumholz2009}, the fraction of H$_2$-to-HI is determined by gas surface density and metallicity in ISM. In BR prescription \citep*{Blitz2006}, the fraction of H$_2$-to-HI is related to the interstellar pressure (see detail in Section 2 in \citealt{Fu2013}). Based on the model outputs, the mock catalogues are created with the light-cone algorithm by \cite{Blaizot2005} and emission line profiles by \cite{Obreschkow2009} for gas distribution on galaxy disks.

\begin{figure*}
\centering
  \includegraphics[angle=270,scale=0.7]{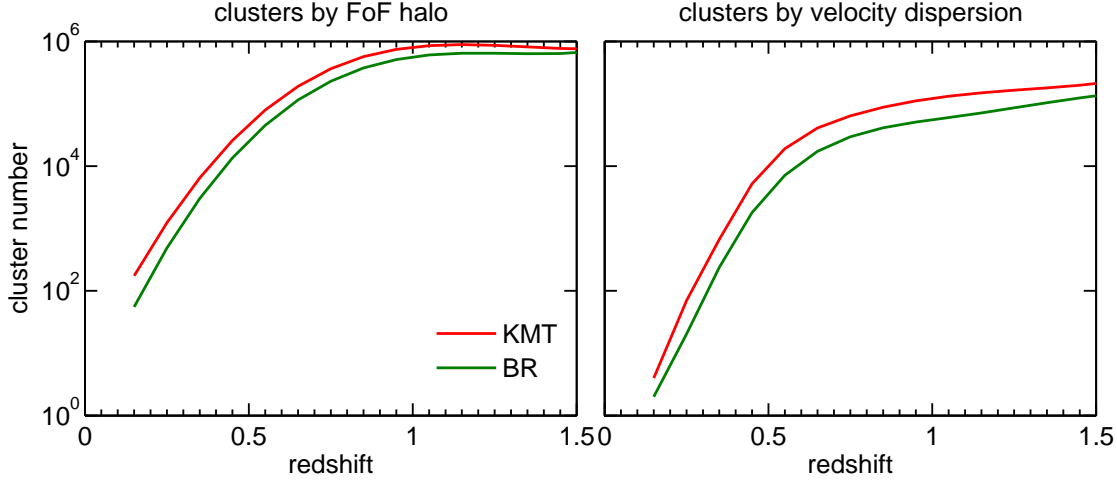}
  \caption{The redshift evolution of galaxy cluster number with $M_{\rm HI} > 10^{11}M_{\odot}$ based on the outputs of L-Galaxies semi-analytic models in FAST observable sky region. The left panel represents the results in which clusters are defined by halo merger history and the right panel show the results in which clusters are defined by velocity and redshift difference. In each panel, the red and green curves represent model results from KMT prescription and BR prescription respectively.}
\end{figure*}

\begin{table*}
\centering
\caption[]{The number of galaxy clusters in different redshift bins shown in Fig.4}
 \begin{tabular}{ccccccccccccc}
  \hline\noalign{\smallskip}
Redshift& 0-0.3   &0.3-0.6&0.6-0.9 &0.9-1.2  &1.2-1.5  \\
  \hline\noalign{\smallskip}
  &&$10^3$&$10^4$&$10^5$&$10^5$\\
  \hline
KMT (FoF)  &    175 &   26&     36&    8.2   &  8.6\\
BR (FoF)    &   55   &  14&     23&    6.1&     6.3\\
KMT (velocity)&   4   &   5.2&  6.4&    1.32&    1.8\\
BR (veolocity) &  2    &   1.8&  2.96&    0.6 &   1.04\\
 \noalign{\smallskip}\hline
\end{tabular}
\end{table*}

\begin{figure}
\centering
 \subfigure{
 \begin{minipage}{6cm}
 \centering
  \includegraphics[angle=270,scale=0.55]{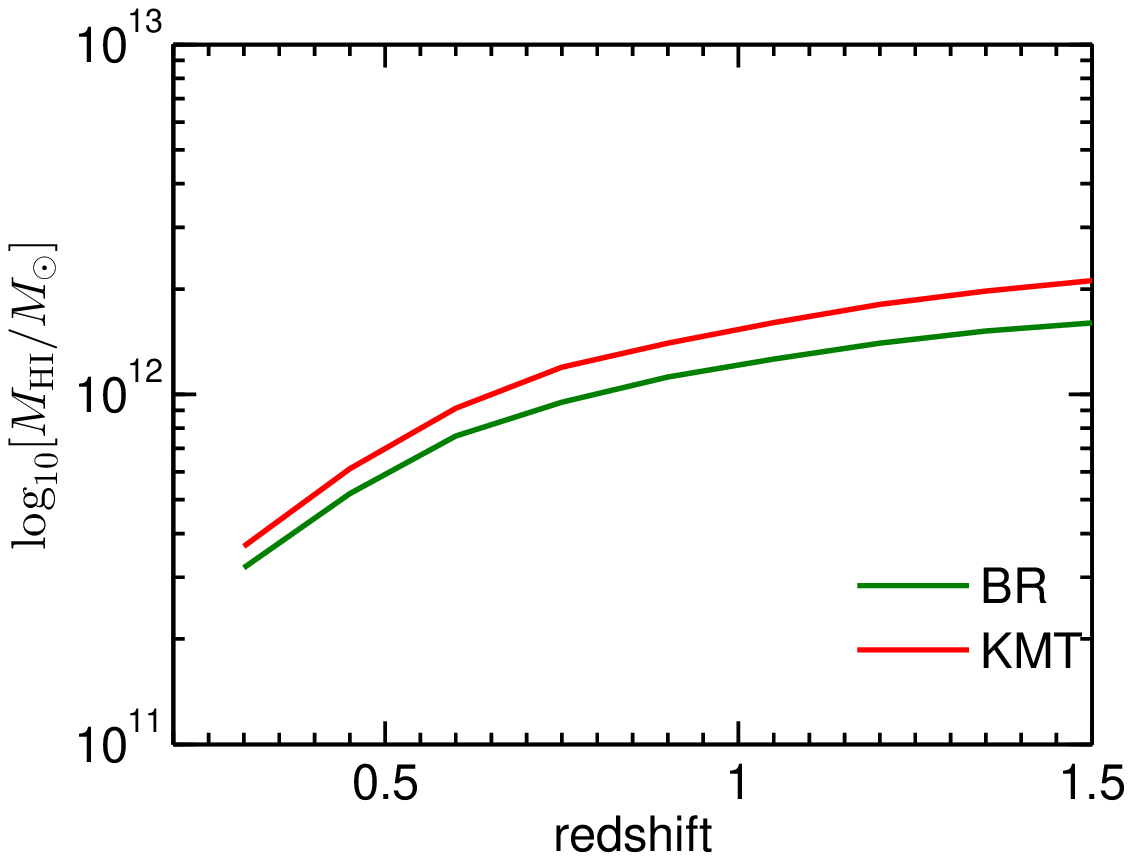}
\end{minipage}
}
\subfigure{
\centering
 \begin{minipage}{6cm}
 \centering
  \includegraphics[angle=270,scale=0.75]{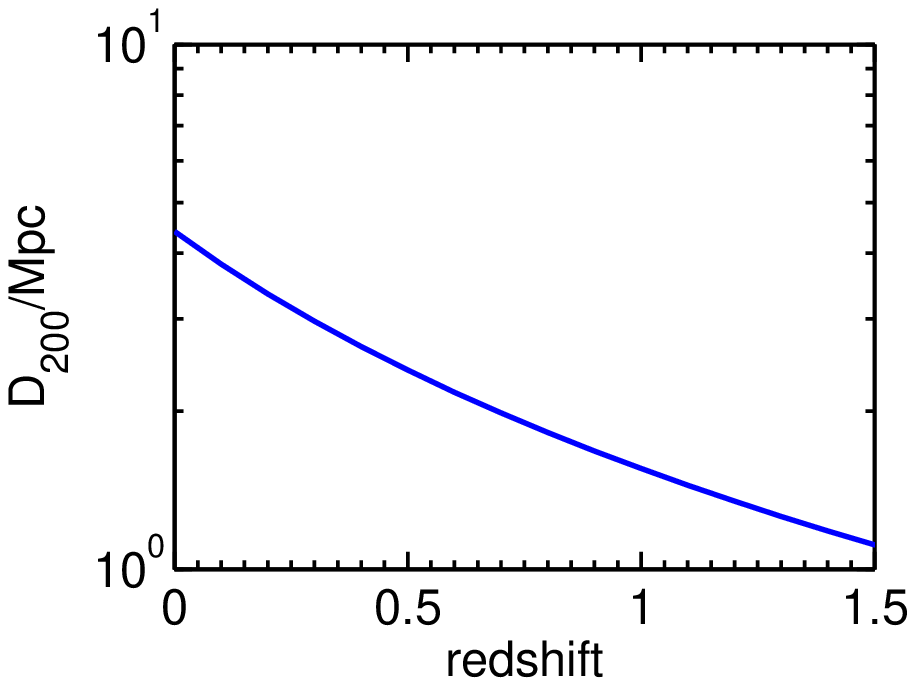}
\end{minipage}
}
\caption{The left panel is the redshift evolution of HI mass in the Virgo size cluster, in which the cluster is defined by the velocity difference. Two curves represent the results of two H2-to-HI transition prescriptions in the models. The right panel is the redshift evolution of the Virgo-like cluster size, in which the cluster is still defined by the velocity dispersion to the central.
Because of the simple relation in the halo evolution, $r_{\rm vir}=0.1H_0^{-1}(1+z)^{-3/2}v_c$, the redshift evolution between cluster size $r_{200}$ and a fixed cluster velocity $v_c$ is approximately proportional to $(1+z)^{-3/2}$ no matter what kind of physical prescription is used}
\end{figure}

Because of the resolution limit of FAST, it is only possible to detect the integrated HI emission at high redshifts in galaxy groups or clusters. In the mock catalogues, two methods are used to get the clustering properties of galaxies. In the first method, we define the galaxy clusters according to the merger history in the dark matter simulation, i.e the galaxies in the same subhalo are treated as a galaxy cluster. In the second method, we define the galaxy clusters similar to the observations, i.e the galaxies are in the same cluster if the velocity difference is lower than 1500km/s or the redshift difference is lower than 0.005. Considering the beam size of FAST, we only add the HI mass within a galaxy cluster covered by the FAST beam of $2.95(1+z)$~arcmin.

According to the detection limit of FAST for HI gas at redshift $z>0.4$ in Table 1, we give the predictions for the number of groups or clusters in Fig.4. The clusters are selected with $M_{\rm HI} > 10^{11}M_{\odot}$ in the sky region with declination from -10 $^\circ$ to 60 $^\circ$ at redshift 0.3-1.5. As discussed above, the left and right panels represent the results of two different definitions of galaxy clusters in the models. In the right panel, the cluster numbers are a lot lower than the left panel when we use the velocity and redshift difference to define a galaxy cluster, which means many galaxies in different subhalos based on the halo merger history may be treated as one galaxy cluster or group in observations because of the projection effect and relatively low velocity difference for galaxies in dense environment. In each panel, the red and green curves represent the model results with two different H$_2$-to-HI gas transition prescriptions. The model with KMT prescription predicts more HI gas rich clusters than that of the BR prescription, which is consistent with the result that KMT prescription gives higher HI-to-H$_2$ at high redshift in previous model work \citep{Fu2012}.

For the results at $z<0.3$, models predict very few gas rich clusters with HI gas mass over
$10^{11}M_{\odot}$. The KMT prescription predicts 4 clusters and BR prescription predicts only two, which should be still a bit higher than the real universe, since Virgo is the only gas rich cluster in nearby universe. On the other hand, the models show that the cluster number increases a lot to about $10^5$ at $z>1$, which indicates that star formations and supernova feedbacks consume a large fraction of gas at $0.3<z<1$.

Fig.5 shows the redshift evolution of the size and HI mass of a Virgo-like cluster based on the SAM simulations. We can see that the HI mass increases by a factor of about 2-3  and the size decreases by a factor of about 2-3 from redshift z=0.3 to z=1. Hence we expect that much more HI fluxes will be included in the FAST beam at high redshift and the integration time listed in Table 1 could be reduced by a factor of 4-9.

\section{Observation strategy and sample selection for future FAST HI studies}

\subsection{Targeted observations of selected gas rich clusters}

Although FAST has very high sensitivity, it still needs hours to detect a high z group or cluster. To make good use of telescope time, careful selection of targets are critical. We would select clusters that have the following properties:\\
a. optically selected clusters or groups  that have blue colors;\\
b. located in intermidiate high density environment. \\
The second criterion is based on the fact that on large scale, HI follows the dark matter distribution and more HI gas are expected along the cosmic web structures.

It has long been known that galaxy properties such as optical color,  morphology and star formation rates are closely correlated with the galaxy environment which is measured by the galaxy number density.
For example, \cite{Dressler1980} pointed out that the fraction of early type and S0 galaxies increase with increasing density environment. \cite{Balogh1997} also found that the fraction of star forming galaxies is smaller in cluster environment than in the field.
Since spiral galaxies are normally gas rich, we expect that the HI gas fractions in blue star forming galaxies are high. Thus
we need to select clusters dominated by star-forming galaxies
which are likely to (still) contain significant HI masses, not having collapsed sufficiently to the state
where ram-pressure stripping and dynamical interactions induce the HI deficiency seen in massive
low redshift clusters like Coma.  For example,  \cite{Rakos1995}  have compiled a sample of 17 clusters in a study of B-O effect ,
 with redshifts range from 0 to 0.9.  In this sample,  the blue galaxy fraction  increases from  20\%  at z=0.2 to 80\% at z=0.9.
Clusters such as CL1322.5+3027(z=0.750), CL1622.5+2352(=0.927)
 with a high fraction of blue galaxies could be good targets for future FAST deep HI observations.

Most clusters and big groups have $M_{\rm HI}/M_{\rm v}=10^{-3}$ to $10^{-4}$, thus
for a Virgo type clusters, which has $M_{\rm v}=10^{15} M_{\odot}$, we will have
$M_{\rm HI}=10^{11} M_{\odot}$ to $10^{12} M_{\odot}$. The extreme systems with $M_{\rm HI}=10^{12} M_{\odot}$, such as those shown in Fig. 5,
should be detectable by FAST at z $\sim$1, in 30-60 minutes.

For clusters with $z < 0.35$, the 19 beam array receiver can be used to scan the whole cluster region. When $z > 0.35$, only the $\rm 560MHz-1120$ MHz single beam receiver is available at FAST. The system temperature of this receiver is expected to be about 20\%-30\% better than that of the 19-beam receiver, thus the integration time could be reduced by about 40\% comparing to the values listed in Table 1. Clusters with $z$ greater than 0.6-0.7 can be completely covered by the FAST beam, thus the single-beam receiver would be the best choice for observing high z clusters.

\subsection{Blind sky surveys}

As predicted  by numerical simulations, a blind all sky HI survey using the 19 beam array receiver of FAST can detect large amount of groups and clusters with redshifts range from 0 to 0.35. Using the drift scan mode (Qian et al.2017 in preparation), and integrate 20 s per point, we can cover the sky area of $\rm 0^h<R.A.<24^{h},-10^{\circ}<decl<60^{\circ}$  in one or two years.  However, deeper surveys are required to detect the major portion of member galaxies in clusters with $z > $0.4. If we choose 120s of integration time per point,  it could take more than 10 years to survey the full FAST sky. Thus it is more practical to select particular regions for a deep blind sky survey.  Higher priority should be given to regions that have  high sensitivity data in optical wavelengths. The SDSS survey could miss many optically week but HI bright galaxies at redshift $z > $0.04,  as shown in section 3.1.3. Future surveys from
the LSST may be able to provide data complementary to the FAST HI survey.

Multibeam receiver is also useful for searching HI in clusters over a large part of sky at intermediate redshifts. For example, if a seven-beam array receiver operating at 500-1000 MHz can be developed for FAST,  it would become feasible to make deep blind survey of HI clusters.  SAM simulations predict that there are about $10^4-10^5$ clusters in the sky region of $\rm -10^{\circ}<decl<60^{\circ}$.  Even if we blindly scan  1\% of the sky region repeatedly, with a typical integration time of 30-60 minutes per point, we could detect about 100-1000 HI clusters with $z> 0.5$, and significantly improve the
sample for studying the HI evolution effect in clusters.

\section{Summary and conclusion}

In summary, we have used HI data from the ALFALFA survey to derive the integrated HI emission of the Virgo cluster.  We further use Virgo as an example to evalute the feasiblity of using FAST to detect the integrated HI emission at high redshifts.  We have estimated the  integration time to detect a Virgo type cluster at different redshifts. At redshift of 0.7, a Virgo sized cluster could be completely covered by the FAST beam and the integration time to detect such a cluster at this redshift by FAST is about 5 hours, assuming no evolutionary effect for the HI contents. We have also employed a SAM model to simulate  the evolution of HI contents in galaxy clusters.  Our simulations suggest that the HI mass of a Virgo-like cluster could be 2-3 times higher  and the physical size could be more than 50\% smaller from z=0.3 to z=1. Thus the integration time could be reduced by a factor of 4-9 for cluster at $z \sim 1$. The well known Butcher-Oemler effect also suggests that clusters at higher redshifts contain more spiral galaxies than that of local clusters. Thus we conclude that it is feasible for FAST to detect the integrated HI emission at redshifts around 1.
Our SAM simulations suggest that there are $10^4$ - $10^5$ clusters with a HI mass greater than 10$^{11} M_{\odot}$ in the sky region visible to FAST.

For the local universe,  we have also used SAM simulation to create mock catalogs of clusters to predict the outcomes from the FAST all sky surveys.
Comparing with the optically selected catalogs derived by
cross matching the galaxy catalogs from the SDSS survey and the ALFALFA survey,  we find that the HI mass distribution of the mock catalog with 20 second integration time agrees well with that of observations.  However, the mock catalog with 120 second integration time predicts much more groups and clusters that contains a  population of low mass HI galaxies not detected by the ALFALFA survey.  Future deep HI blind sky surveys with FAST would be able to test such prediction and set constraints to the numerical simulation models.

\normalem
\begin{acknowledgements}

We thank Dr. Alan Duffy for providing the mock catalog for FAST, Dr. Rurong Chen for helps with analyzing the catalog, Dr. Zhonglue Wen for providing us his WHL cluster member galaxies.  We also thanks Dr. Yu, Gao, Dr. Mathra Haynes and Dr. Riccardo Giovanelli for valuable suggestions to this project.  MZ acknowledges the support by NSFC grant no. U1531246, and by the China Ministry of Science and Technology under the State Key Development Program for Basic Research (2012CB821800).  Jian Fu acknowledges the support by NSFC no. U1531123, the Youth innovation Promotion Association CAS, and the Opening Project of Key Laboratory of Computational Astrophysics, National Astronomical Observatories, CAS. We acknowledge the work of the entire ALFALFA collaboration team in observing, flagging, and extracting the catalog of galaxies used in this work. The Arecibo Observatory is operated by SRI International under a cooperative agreement with the National Science Foundation (AST-1100968), and in alliance with Ana G. Mndez-Universidad Metropolitana, and the Universities Space Research Association. Funding for SDSS-III has been provided by the Alfred P. Sloan Foundation, the Participating Institutions, the National Science Foundation and the US Department of Energy Office of Science.

\end{acknowledgements}

\bibliographystyle{raa}

\end{document}